\documentclass[aps,prl,reprint,showpacs]{revtex4-1}
\usepackage[dvipdfmx]{graphicx}
\usepackage{amsmath,amssymb}
\usepackage{color}
\usepackage{bm}
\usepackage{ulem}

\usepackage{braket}

\begin{document}

\title{Sub-band picture of high-harmonic generation in solids}
\author{Tomohiro Tamaya}
\email{tamaya@issp.u-tokyo.ac.jp}
\author{Takeo Kato}
\affiliation{
${}$Institute for Solid State Physics, University of Tokyo, Kashiwa, 277-8581, Japan}

\date{\today}

\begin{abstract}
We propose a novel picture of high-harmonic generation (HHG) in solids based on the concept of temporally changing band structures. 
To demonstrate the utility of this picture, we focus on the high-order sideband generation (HSG) caused by strong terahertz (THz) and weak near-infrared (NIR) light in the context of pump-probe spectroscopy. 
We find that the NIR frequency dependence of the HSG indicates the existence of new energy levels (sub-bands) around the band-gap energy, which have multiple frequencies of THz light. 
This sub-band picture explains why the HSG intensity becomes a non-monotonic function of the THz light amplitude.
The present analysis not only reveals the origin of the plateau structure in HHG spectra, but also provides a connection to other high-field phenomena.
\end{abstract}

\maketitle
High-harmonic generation (HHG) is one of the most fundamental topics of nonlinear optics~\cite{Shen1984,Boyd1992,Yariv1984}. 
In particular, HHG in gaseous media has made it possible to develop a high-frequency light source and has paved the way for attosecond science~\cite{Protopapas1997,Brabec2000,Agostini2004,Krausz2009,Corkum2007,Corkum1993}. 
In recent years, HHG in solids {has} been experimentally observed, and its diversity has led to a new research field and expectations of novel optical devices~\cite{Ghimire2011,Schubert2014,Luu2015,Vampa2015,Hohenleutner2015,Liu2017,You2017,Langer2017,Tamaya2017Science,Kaneshima2018,Vampa2018,Ndabashimiye2018,Langer2018,Kaneshima2018,Silva2018,Saito2017}.
In fact, HHG has been explored in a wide variety of solids, and by clarifying the universal properties of HHG in solids, much progress in high-intensity optical technology can be expected.

%The HHG mechanism in the atomic and solid-state cases {has been} understood in terms of a three-step model that makes an analogy between tunnel ionization in gaseous media and Zener tunneling in semiconductors~\cite{Tamaya2016PRL,Tamaya2016PRBR,Tamaya2017Science}. 
The HHG mechanism in solids has been mainly explained in terms of either the three-step model or the Bloch oscillation~\cite{Ghimire2011,Schubert2014,Luu2015,Vampa2015,Hohenleutner2015,Liu2017,You2017,Langer2017,Tamaya2017Science,Kaneshima2018,Vampa2018,Ndabashimiye2018,Langer2018,Kaneshima2018,Silva2018,Saito2017,Tamaya2016PRL,Tamaya2016PRBR,Tamaya2017Science,Tao2018,Xiao2019,McDonald2015}.
%In the three-step model \cite{Protopapas1997,Brabec2000,Agostini2004,Krausz2009,Corkum2007,Corkum1993}, however, the connection between HHG and other high-field phenomena, such as the dynamical Franz-Keldysh effect \cite{Nordstrom1997,Nordstrom1998,Srivastava2004,Jauho1996}, above-threshold ionization~\cite{Agostini1979,Cormier1997,Eberly1991}, and coherent destruction of tunneling~\cite{Grossmann1991,Lignier2007,Platero2004}, is unclear. 
In these models, however, the connection between HHG and other high-field phenomena, such as the dynamical Franz-Keldysh effect \cite{Nordstrom1997,Nordstrom1998,Srivastava2004,Jauho1996,Lucchini2016}, above-threshold ionization~\cite{Agostini1979,Cormier1997,Eberly1991}, and coherent destruction of tunneling~\cite{Grossmann1991,Lignier2007,Platero2004}, is unclear.
%It appears that HHG and these optical phenomena are different aspects of the same electron interaction between the materials and light field, and revealing their connections would provide new aspects of HHG. 
It appears that HHG and these optical phenomena are different aspects of the same electron interaction with the strong light field, and revealing their connections would provide new aspects of HHG. 
Moreover, clarifying them might lay the foundation for understanding high-intensity optical phenomena and reveal a different HHG mechanism in solids.

A representative method of exploring the HHG mechanism is pump-probe spectroscopy, where weak near-infrared {(NIR)} and strong terahertz (THz) light are simultaneously imposed.
Recent studies have claimed that the two-color light yields high-order sideband generation (HSG)~\cite{Kono1997,Zaks2013,Liu2018OptLett,Langer2016,Zaks2012,Cerne1997,Wagner2011,Luu2018,Yan2017} whose properties are determined by quasiparticle collisions~\cite{Langer2016,Zaks2012}; i.e., carriers excited by the weak NIR light are driven by the strong THz light, after which they collide and induce HSG. 
This physical interpretation is similar to the one for gaseous media~\cite{Jin2014,Ishikawa2003,Takahashi2007,Banks2013,Zeng2002}. 
HSG experiments, however, can be understood in a different way, i.e., as a probe of the {\it modified} states of solids under strong THz light irradiation, assisted by one-photon excitation due to the weak NIR light. 
Therefore, tuning the NIR frequency as well as the intensity of the THz light would be a pump-probe spectroscopy for the HHG mechanism and changes in the HSG spectra should be important clues to the non-perturbative mechanisms of HHG.

In this paper, we theoretically investigate how the properties of HSG in solids change depending on the THz intensity and NIR frequency. 
Our numerical results suggest a sub-band picture of HHG in solids that indicates a new aspect of HHG and clarifies the connection {to} other high-field phenomena.

\begin{figure*}[tb]
\begin{center}
\includegraphics[width=17cm]{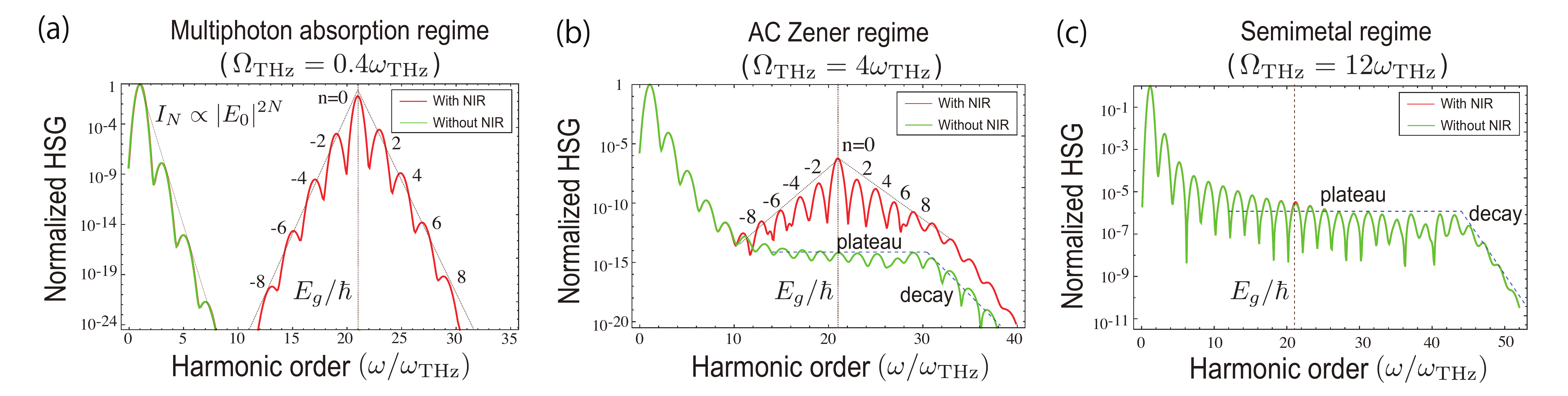}
\caption{(Color online) Normalized high-harmonic spectra caused by high-intensity THz and low-intensity NIR light {for} (a) multiphoton absorption regime $(\Omega_{\rm THz}=0.4\omega_{\rm THz})$, (b) AC Zener regime $(\Omega_{\rm THz}=4\omega_{\rm THz})$, and (c) semimetal regime $(\Omega_{\rm THz}=12\omega_{\rm THz})$. 
{The band-gap energy is set to $E_{g}=21\hbar \omega_{\rm THz}$.} The green and red lines indicate HHG (without {the} NIR light) and HSG spectra (with {the} NIR light), respectively.}
\label{fig:HSG_spectra}
\end{center}
\end{figure*}

To investigate the properties of HSG, we {have extended} the previous theories~\cite{Tamaya2016PRL,Tamaya2016PRBR,Tamaya2017Science} to the case of a two-color pump-probe system. 
In this work, we consider a two-dimensional covalent crystal with two different atoms in a unit cell as a minimum model in which space-inversion symmetry is kept. 
The theoretical model employed here is simple, but has enough facility to clarify the essence of HSG. 
By performing a similar procedure to that of Ref.~\cite{Tamaya2016PRL,Tamaya2016PRBR,Tamaya2017Science}, we arrive at {the} Hamiltonian $H=H_{0}+{H_{\rm intra}+H_{\rm intra}}$, where~\cite{supplemental}
\begin{eqnarray}
&&H_{0}=\sum_{\bm{k}}\left[E_{c}\left(\bm{k}\right)a^{\dagger}_{c\bm{k}}a_{c\bm{k}}+E_{v}\left(\bm{k}\right)a^{\dagger}_{v\bm{k}}a_{v\bm{k}}\right], \label{eq:Hamiltonian0} \\ 
&&H_{{\rm intra}}= \sum_{\bm{k},\nu}{\hbar \Omega_{\nu}(t)} \cos\theta_{\bm{k}}\left(a^{\dagger}_{c\bm{k}}a_{c\bm{k}}-a^{\dagger}_{v\bm{k}}a_{v\bm{k}}\right) ,
 \label{eq:Hamiltonian1}\\
&&H_{{\rm inter}}= -\sum_{\bm{k},\nu}{\hbar \Omega_{\nu}(t)} \sin\theta_{\bm{k}}\left(a^{\dagger}_{c\bm{k}}a_{v\bm{k}}+a^{\dagger}_{v\bm{k}}a_{c\bm{k}}\right).
 \label{eq:Hamiltonian2}
\end{eqnarray}
The Hamiltonian $H_0$ describes a bulk semiconductor, where $a_{c\bm{k}}$ ($a_{v\bm{k}}$) is the annihilation operator of conduction (valence) electrons, and the dispersion is given as $E_{c(v)}=\hbar^2 \bm{k}^2/2m_{c(v)} \pm E_{g}/{2}$ ($\hbar$ is the reduced Planck constant, $\bm{k}$ is the two-dimensional Bloch wavevector, $m_{c(v)}$ is the effective mass of the conduction (valence) band, and $E_{g}$ is the band-gap energy). 
In this paper, we suppose $m_{c}=-m_{v}$, i.e., $E_{c}(\bm{k})=-E_{v}(\bm{k})=E(\bm{k})$. 
The light-matter interaction is expressed by $H_{\rm inter}$ and $H_{\rm intra}$, where the former (latter) describes the interband (intraband) transition. 
It is remarkable that the intraband Hamiltonian $H_{\rm intra}$ can be renormalized into a bulk Hamiltonian $H_{0}$, which leads us to the idea of a temporally changing band structure defined by $\xi_{c(v)}{({\bm k},t)}=E_{c(v)}(\bm{k}) \pm \hbar \cos\theta_{\bm{k}} \sum_{\nu} \Omega_{\nu}(t)${~\cite{Tamaya2016PRL,Tamaya2016PRBR,Tamaya2017Science}}. 
Below, we assume that the Rabi frequency $\Omega_\nu(t)$ ($\nu = {\rm NIR \ or \ THz}$) has the following time profile:
\begin{align}
 {\Omega_{\nu}(t)=\Omega_\nu \exp\left[-(t-t_{0})^2/T^2 \right] \cos \omega_\nu t},
\end{align}
where $\Omega_\nu$ and $\omega_\nu$ are the maximum amplitudes of the Rabi frequency and the incident frequency of light, respectively \cite{Haug2009}. 
Throughout this paper, the parameters of the incident {light} will be fixed at $t_{0}=24\pi/\omega_{\rm THz}$ and $T=4\pi/\omega_{\rm THz}$.

Using the above Hamiltonian, the time-evolution equations for the wavefunction ${\psi}_{c\bm{k}}$ (${\psi}_{v\bm{k}}$) of the conduction (valence) electrons can be derived as \cite{supplemental}
\begin{eqnarray}
 i \hbar \dot{{\psi}}_{c\bm{k}}&=&\xi_{c}(\bm{k},t){\psi}_{c\bm{k}}- \sin\theta_{\bm{k}}\sum_{\nu}{\hbar \Omega_{\nu}(t)} {\psi}_{v\bm{k}}, \label{eq:kinc} \\
 i \hbar \dot{{\psi}}_{v\bm{k}}&=&\xi_{v}(\bm{k},t){\psi}_{v\bm{k}}-\sin\theta_{\bm{k}}\sum_{\nu}{\hbar \Omega_{\nu}(t)} {\psi}_{c\bm{k}}, \label{eq:kinv}
\end{eqnarray} 
with the initial conditions, $\psi_{c{\bm k}}(0)=0$ and $\psi_{v{\bm k}}(0)=1$. 
%\del{Here, $\xi_{c(v)}{({\bm k},t)}$ describes a time-dependent band shift due to $H_{\rm intra}$, which will be crucial to the following analysis.}
The numerical solutions of Eqs.~(\ref{eq:kinc}) and (\ref{eq:kinv}) give the time evolution of ${\psi}_{c\bm{k}}$ and ${\psi}_{v\bm{k}}$. 
The temporal variations of the generated currents can be calculated using the definition, ${\cal J}(t)=-c\sum_{\nu}\langle{\partial H_{I}}/{\partial A_{\nu}(t)}\rangle \propto \sum_{\bm{k}}[\cos\theta_{\bm{k}} (|{\psi}_{c\bm{k}}|^2-|{\psi}_{v\bm{k}}|^2) -\sin\theta_{\bm{k}}({\psi}^{*}_{v\bm{k}} {\psi}_{c\bm{k}}+{\psi}^{*}_{c\bm{k}}{\psi}_{v\bm{k}})]$. 
We can derive the HSG intensity spectra as $I=\left|\omega {\cal J}(\omega) \right|^2$, where ${\cal J}(\omega)$ is the Fourier transform of the generated current. 
Below, we set the band-gap energy of the semiconductor as $E_{g}=21\hbar \omega_{\rm THz}$ and discuss the difference between the HHG ($\Omega_{\rm NIR}=0$) and the HSG ($\Omega_{\rm NIR}=10^{-4}\omega_{\rm THz}$) spectra.

\begin{figure*}[tb]
\begin{center}
\includegraphics[width=17cm]{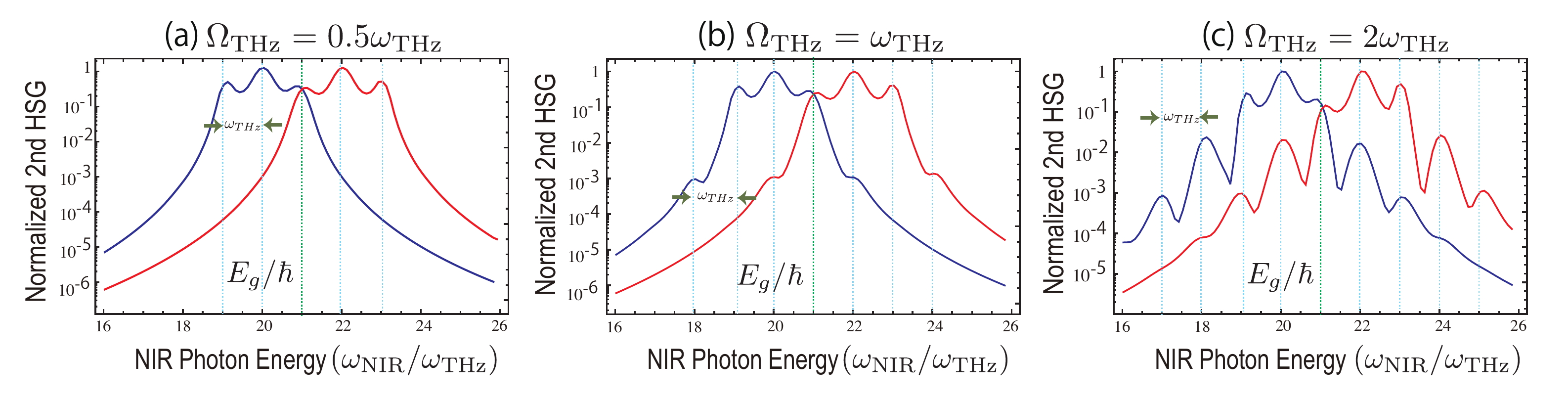}
\caption{(Color online) NIR frequency dependence of $\pm$2nd HSG for $E_{g}=21 \hbar \omega_{\rm THz}$ in the case of (a) $\Omega_{\rm THz}=0.5\omega_{\rm THz}$, (b) $\Omega_{\rm THz}=\omega_{\rm THz}$, and (c) $\Omega_{\rm THz}=2\omega_{\rm THz}$. 
The blue and red lines indicate positive and negative 2nd HSG, respectively.}
\label{fig:Frequency_dependence}
\end{center}
\end{figure*}

First, let us consider the resonant case of $\omega_{\rm NIR}=E_g/\hbar=21\omega_{\rm THz}$.
The numerically calculated HHG and HSG spectra are shown in Fig.~\ref{fig:HSG_spectra} for the (a) multiphoton absorption regime $({\Omega_{\rm THz}}=0.4\omega_{\rm THz})$, (b) AC Zener regime $({\Omega_{\rm THz}}=4\omega_{\rm THz})$, and (c) semimetal regime $({\Omega_{\rm THz}}=12\omega_{\rm THz})$, respectively~\cite{Tamaya2016PRL,Tamaya2016PRBR,Tamaya2017Science}. Here, the green and red lines indicate the HHG and HSG spectra. 
In the multiphoton absorption regime (Fig.~\ref{fig:HSG_spectra}(a)), we find that the low-order harmonics (1st-7th) indicate the well-known relation of conventional nonlinear optics, wherein $I_{N} \propto |P_{N}|^2 \propto |E_{0}|^{2N}$ for both the HHS and HSG spectra. 
For the HSG spectra, we also find a peak at $\omega = \omega_{\rm NIR}$ (denoted with $n=0$) accompanied with side peaks at $\omega = \omega_{\rm NIR} + n\omega_{\rm THz}(\equiv \omega_{{\rm HSG},n})$ ($n=\pm 2, \pm 4, \cdots$), whose intensity is proportional to $|E_{0}|^{2|n|}$. 
Hereafter, the difference between the HHG and HSG spectra is called the HSG signal. 
In the AC Zener regime (Fig.~\ref{fig:HSG_spectra}(b)), plateau and decay regions appear in the HHG spectra. 
Although the peaks at $\omega = \omega_{{\rm HSG},n}$ still remain in the HSG spectra, their intensities are suppressed, and approach those for HHG with increasing $\Omega_{\rm THz}$. 
While the HSG signal decays away from the main peak at $\omega = \omega_{\rm NIR}$, its decay rate is more moderate than that for the multiphoton absorption regime. 
In the semimetal regime (Fig.~\ref{fig:HSG_spectra}(c)), the HSG and HHG spectra nearly coincide, and a clear HSG signal does not appear.

These features in the HSG spectra can be understood by considering a temporally changing band structure denoted with $\xi_{c(v)}({\bm k},t)$, which originates from the intraband transition described by $H_{\rm intra}${~\cite{Tamaya2016PRL,Tamaya2016PRBR,Tamaya2017Science}}. 
In the multiphoton absorption regime, the side peak of $n>0$ ($n<0$) in the HSG is just the sum (difference) frequency generation between the THz and NIR lights~\cite{Shen1984,Boyd1992,Yariv1984}. 
With increasing $\Omega_{\rm THz}$, the temporal shift of the band structure disturbs the complete resonant condition, $\omega_{\rm NIR}=E_g/\hbar$, and results in a decrease of the HSG signal. 
Upon further increase of $\Omega_{\rm THz}$, the conduction and valence bands start to overlap, and the system is driven into a semimetal state, for which the HSG signal disappears, as the resonant condition makes no sense there.

The present HSG signal can be regarded as the result of pump-probe spectroscopy using the high-intensity THz pump light and the low-intensity NIR probe light. 
From this viewpoint, let us consider the HSG signal while sweeping the NIR frequency in {the} AC Zener regime. 
Figures~\ref{fig:Frequency_dependence}~(a)-(c) plot the side peak intensities of $n=2$ (the blue line) and $n=-2$ (the red line) as a function of $\omega_{\rm NIR}$ for $\Omega_{\rm THz}/\omega_{\rm THz}=0.5$, $1$, and $2$, respectively. 
For $\Omega_{\rm THz}/\omega_{\rm THz}=0.5$ (Fig.~\ref{fig:Frequency_dependence}~(a)), the HSG signal of $n=2$ has the largest maximum when the resonant condition, $\omega_{{\rm HSG},2} = \omega_{\rm NIR} + 2\omega_{\rm THz} = E_g/\hbar + \omega_{\rm THz}$, is satisfied and also has two maxima when $\omega_{{\rm HSG},2}=E_{g}/\hbar$ and $\omega_{{\rm HSG},2}=E_{g}/\hbar +2 \omega_{\rm THz}$. 
Similarly, the HSG signal of $n=-2$ has maxima when $\omega_{{\rm HSG},-2} = E_g/\hbar, E_{g}/\hbar - \omega_{\rm THz}$, and $E_{g}/\hbar - 2\omega_{\rm THz}$. 
With increasing $\Omega_{\rm THz}$, more peaks become visible (Fig.~\ref{fig:Frequency_dependence}(b)), until all are clearly displayed (Fig.~\ref{fig:Frequency_dependence}(c)). 
These peaks at different frequencies of $\omega_{\rm THz}$ in the NIR probe spectra reflect the {\it modified} state of the system under strong THz light irradiation.

\begin{figure}[tb]
\begin{center}
\includegraphics[width=8cm]{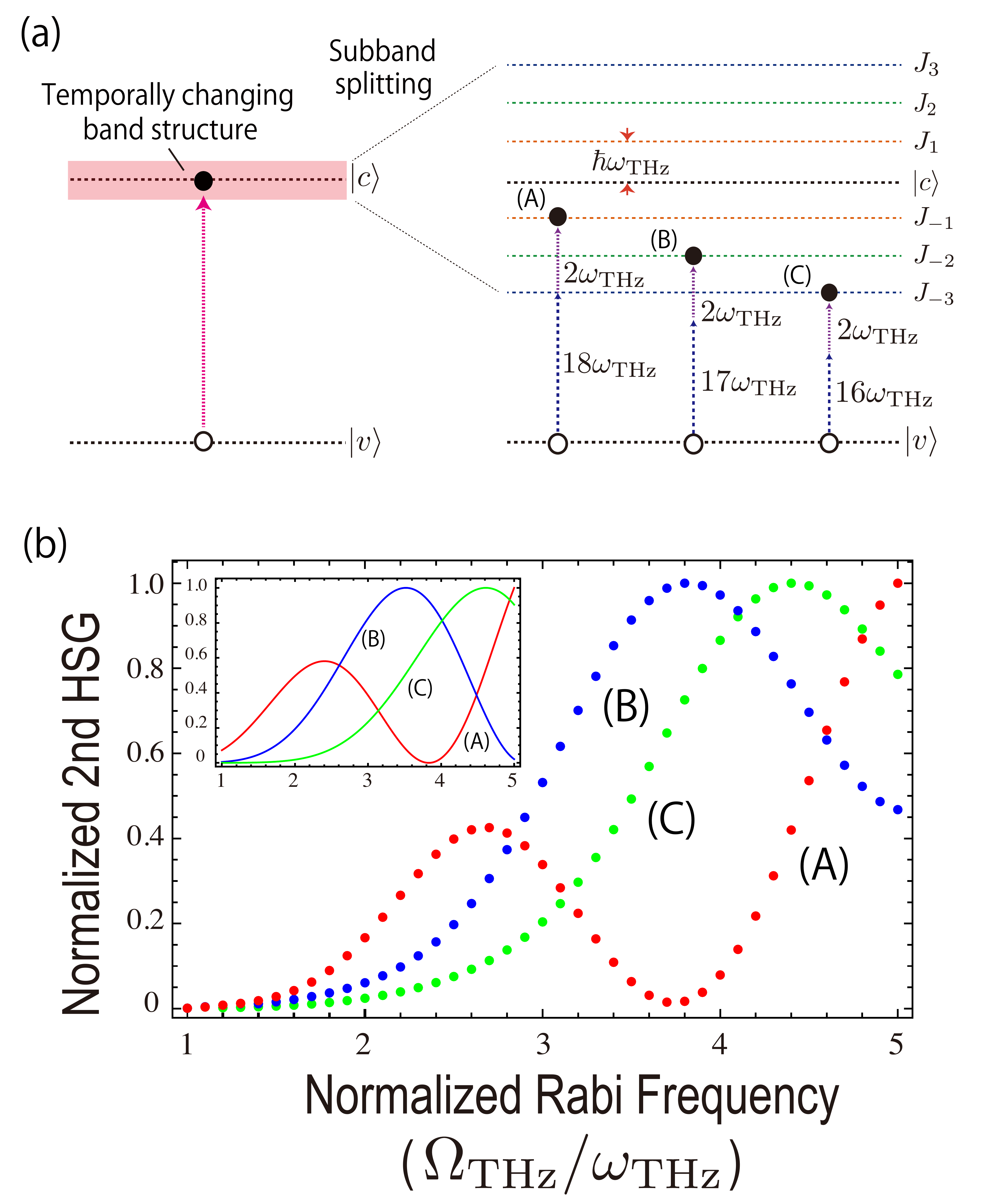}
\caption{(Color online) (a) Schematic diagram of the sub-band picture originating from temporally changing band structure. (b) Numerical results of positive 2nd HSG as a function of $\Omega_{{\rm THz}}/\omega_{{\rm THz}}$. Here, the red, blue, and green dots indicate (A) $\omega_{{\rm NIR}}=18\omega_{{\rm THz}}$, (B) $\omega_{{\rm NIR}}=17\omega_{{\rm THz}}$, and (C) $\omega_{{\rm NIR}}=16\omega_{{\rm THz}}$, respectively, {whose processes are} illustrated in (a). Inset figure: Analytic results of 2nd HSG as a function of $\Omega_{\rm THz}/\omega_{\rm THz}$. The red, blue, and green lines show $|x J_{-1}(x)|^2$, $|x J_{-2}(x)|^2$, and $|x J_{-3}(x)|^2$, respectively, where $x \equiv \Omega_{\rm THz}/\omega_{\rm THz}$.}
\label{fig:Bessel_dependence}
\end{center}
\end{figure}

To investigate the influence of strong THz light, let us consider the dynamics for continuum waves ($\Omega_{\rm THz}(t)= \Omega_{\rm THz} \cos \omega_{\rm THz} t$) in the absence of NIR light. 
The formal solution of $\psi_{c\bm{k}}(t)$ in Eq.~(\ref{eq:kinc}) is written as
\begin{eqnarray}
 {\psi}_{c\bm{k}}(t)&=&\sum_{k=-\infty}^{\infty}i J_{k}\left(\frac{\Omega_{\rm THz}}{\omega_{\rm THz}}\right) \Omega_{\rm THz} \sin\theta_{\bm{k}} \nonumber \\
 & & \hspace{7mm} \times \int_{0}^{t} dt' g(t,t') {\psi}_{v\bm{k}}(t'),
 \label{eq:sidebands}
\end{eqnarray}
where $J_{k}(x)$ is the $k$-th Bessel function, and $g(t,t') = \exp \left[-i\left(E\left(\bm{k}\right) + k \hbar \omega_{\rm THz}\right)(t-t') \right] \cos\omega_{\rm THz}t'$. 
This equation indicates that the conduction band is composed of a superposition of sub-bands which have eigen-energies $E(\bm{k}) + k \hbar \omega_{\rm THz}$ (see the exponential part in $g(t,t')$), where $k$ is an integer. 
We stress that the intraband transition described by $H_{\rm intra}$, that is, the concept of a temporally changing band structure, is crucial to the formation of this sub-band picture (see Fig.~\ref{fig:Bessel_dependence}~(a)). 
Moreover, it is worth noting that the formation of new energy levels under a strong external field has been discussed in a similar way in the study of above-threshold ionization, as indicated in experiments on gaseous media \cite{Eberly1991}.

Let us describe our analysis based on Eq.~(\ref{eq:sidebands}). 
The equation indicates that the prefactor of the integral, $J_{k} (\Omega_{\rm THz}/\omega_{\rm THz})\Omega_{\rm THz}$, effectively describes mixing matrix elements between the $k$th sub-band and the valence band, which can be used for a rough estimate of the transition amplitude. 
To check this idea, we plot in Fig.~\ref{fig:Bessel_dependence}~(b) the numerical solution of the intensity of the second positive HSG peak ($n=2$) as a function of $\Omega_{\rm THz}$. 
Here, the red, blue, and green dots indicate (A) $\omega_{\rm NIR} = 18\omega_{\rm THz}$, (B) $\omega_{\rm NIR} = 17\omega_{\rm THz}$, and (C) $\omega_{\rm NIR} = 16\omega_{\rm THz}$, respectively, where these three frequencies {correspond to HSG signals developed for $\Omega_{\rm THz} \gtrsim 2\omega_{\rm THz}$ (see Fig.~\ref{fig:Frequency_dependence}~(c)), and} satisfy the conditions, $\omega_{{\rm HSG},2}=E_g/\hbar - \omega_{\rm THz}$, $E_g/\hbar - 2\omega_{\rm THz}$, and $E_g/\hbar - 3\omega_{\rm THz}$ (see Fig.~\ref{fig:Bessel_dependence}~(a)). 
The intensity of the HSG signal is a non-monotonic function of $\Omega_{\rm THz}$ and becomes almost zero at finite values of $\Omega_{\rm THz}$ for the case (A). 
For comparison, the inset {of Fig.~\ref{fig:Bessel_dependence}~(b)} plots the analytic results expected from the transition amplitude as a function of $\Omega_{\rm THz}$. 
Here, the red, blue, and green lines show $|x J_{-1}(x)|^2$, $|x J_{-2}(x)|^2$, and $|x J_{-3}(x)|^2$, respectively, where $x=\Omega_{\rm THz}/\omega_{\rm THz}$. 
These figures certainly convince us that the non-monotonic behavior of the positive 2nd HSG reflects the effective transition amplitude from the valence band to the sub-bands of the conduction band~\cite{footnote}.

\begin{figure}[tb]
\begin{center}
\includegraphics[width=8cm]{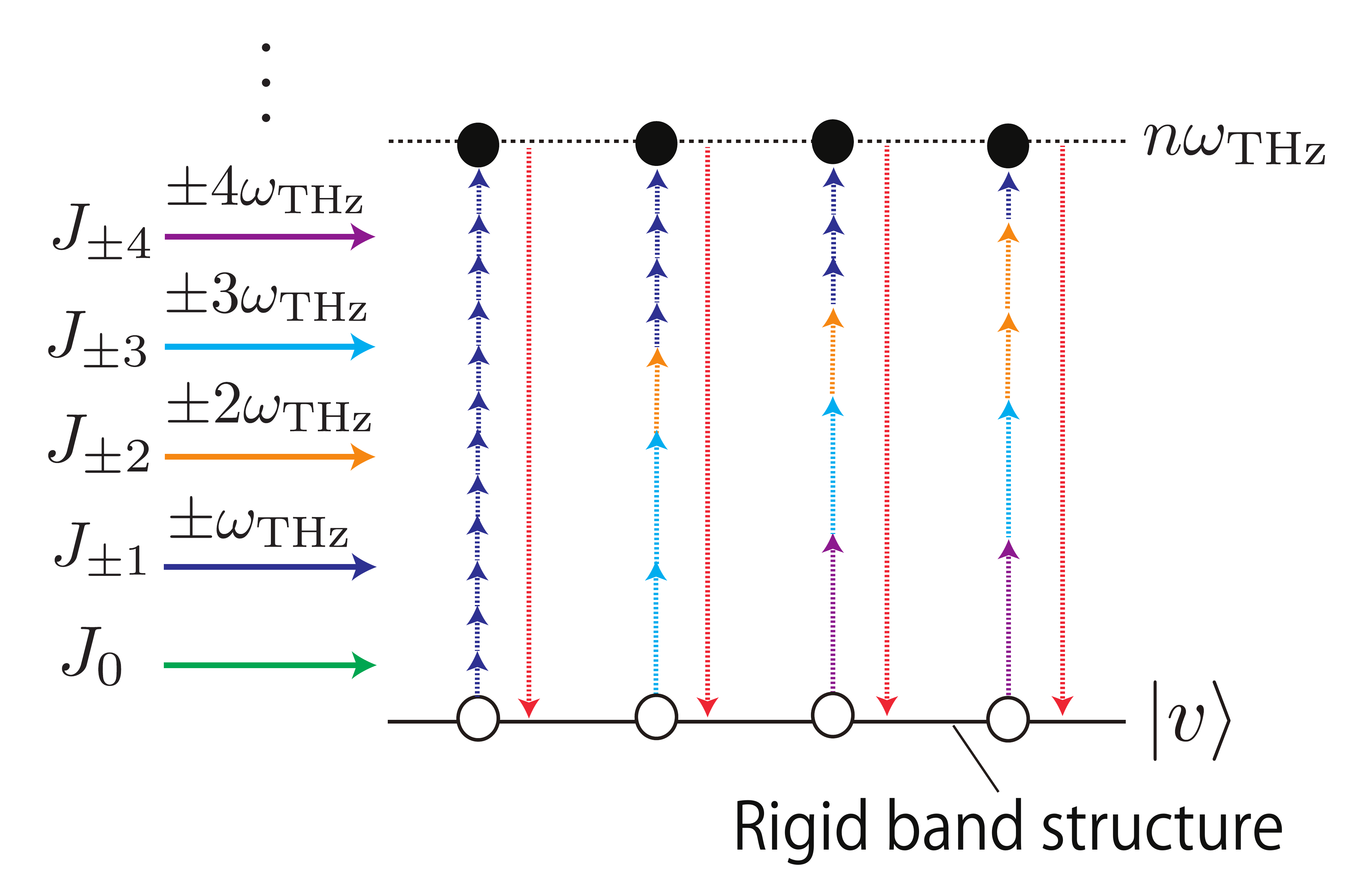}
\caption{(Color online) Schematic diagram of the {excitation paths in the $n$th HHG}. In this picture, the incident THz electric field is modulated as if there exist multiple light sources, whose frequencies and intensities are $n\omega_{\rm THz}$ {($n$ is an integer)} and $J_n(2\Omega_{\rm THz}/\omega_{\rm THz})$, respectively. Various excitation paths constructed from multiple lights are shown by colorful arrows.}
\label{fig:Schematic_Figures}
\end{center}
\end{figure}

The above analysis is useful for gaining a qualitative understanding of the HSG signal.
For a more concrete understanding, we can perform a unitary transformation on the total Hamiltonian $H=H_{0}+H_{\rm{intra}}+H_{\rm{inter}}$ (see Eqs.~(\ref{eq:Hamiltonian0})-(\ref{eq:Hamiltonian2})). 
Here, we introduce a unitary transformation $a_{c\bm{k}}=e^{i\lambda}\sin(\theta_{\bm{k}}/2)\alpha_{c\bm{k}}+e^{-i\lambda}\cos(\theta_{\bm{k}}/2)\alpha_{v\bm{k}}$ and $a_{v\bm{k}}=e^{i\lambda}\cos(\theta_{\bm{k}}/2)\alpha_{c\bm{k}}-e^{-i\lambda}\sin(\theta_{\bm{k}}/2)\alpha_{v\bm{k}}$, where $\lambda= (\Omega_{\rm THz}/\omega_{\rm THz})\sin \omega_{\rm THz}t$ and $\theta_{\bm k}$ is the angle of the wavenumber measured from the $x$-axis. 
By supposing the continuum waves $\Omega_{\rm THz}(t)= \Omega_{\rm THz} \cos \omega_{\rm THz} t$ in the absence of NIR light $\Omega_{\rm NIR}(t)=0$ and performing the above unitary transformation on the Hamiltonian, we can rewrite the Hamiltonian as~\cite{supplemental}
\begin{eqnarray}
H&=&\sum_{\bm{k}}\left[E_{c}(\bm{k})\alpha^{\dagger}_{c\bm{k}}\alpha_{c\bm{k}}+E_{v}(\bm{k})\alpha^{\dagger}_{v\bm{k}}\alpha_{v\bm{k}}\right]\cos\theta_{\bm{k}} \nonumber \\
&+&\sum_{\bm{k},l}E(\bm{k}) \sin\theta_{\bm{k}}J_l \left(\frac{2\Omega_{{\rm THz}}}{\omega_{{\rm THz}}}\right) \nonumber \\
& &\hspace{5mm} \times \left(e^{i l \omega_{{\rm THz}} t}\alpha^{\dagger}_{c\bm{k}}\alpha_{v\bm{k}}+H.c. \right). \label{eq:Hamiltonian20}
\end{eqnarray}
Accordingly, the generated current becomes ${\cal J}(t)=\sum_{\bm{k}}[|\phi_{c\bm{k}}|^2-|\phi_{v\bm{k}}|^2]$, where $\phi_{c\bm{k}}$ and $\phi_{v\bm{k}}$ are the wavefunctions for the new basis. 
This Hamiltonian indicates that the {valence-conduction energy difference} is fixed to {$2E({\bm k})$} (the rigid-band picture), while the incident THz electric fields are modulated as if there are multiple light sources whose frequencies are $l \omega_{{\rm THz}}$ ($l$ is an integer). 
Then, the amplitude of this virtual light source with frequency $l\omega_{\rm THz}$ is proportional to $J_l(2\Omega_{{\rm THZ}}/\omega_{{\rm THz}})$, and therefore, it starts to be effective for $2\Omega_{\rm THZ}/\omega_{\rm THz} \gtrsim l$ because of the nature of the Bessel function.
We should note that the conventional framework of nonlinear optics~\cite{Shen1984,Boyd1992,Yariv1984} is based on the perturbation expansion with respect to light-matter coupling after approximating the Bessel function as $J_0(x)\simeq 1$, $J_1(x) \simeq x$ and neglecting higher-order Bessel functions. 
Next, let us consider the effect of the strong THz light via the $0$th-order Bessel function. 
By diagonalizing the Hamiltonian (\ref{eq:Hamiltonian20}), the eigen-energies are given as $\pm E(\bm{k})[\cos^2\theta_{\bm{k}}+\sin^2\theta_{\bm{k}}J^{2}_{0}(2\Omega_{\rm THz}/\omega_{\rm THz})]^{1/2}$. 
This expression indicates that the energy difference between the conduction and valence bands is suppressed with increasing $\Omega_{\rm THz}$. 
This result suggests the dynamical Franz-Keldysh effect~\cite{Nordstrom1997,Nordstrom1998,Srivastava2004,Jauho1996,Lucchini2016}, though the suppression also depends on the angle $\theta_{\bm k}$. 
The condition for the strongest suppression of the band gap, $J_{0}(2\Omega_{\rm THz}/\omega_{\rm THz})=0$, is in common to that of coherent destruction of tunneling (or dynamical localization) in transport theory of strongly driven systems~\cite{Grossmann1991,Lignier2007,Platero2004}.

Let us discuss the origin of the plateau structure of the HHG from the viewpoint of these multiple light sources. 
By regarding all the light sources with frequencies $l\omega_{THz}$ as perturbations, we can express the HHG spectra by a series expansion with respect to the amplitudes of the light sources. 
The HHG spectra are then determined by the multi-variable polynomials of the amplitudes of the light sources, which are intuitively expressed as excitation paths constructed from multiple lights (see Fig.~\ref{fig:Schematic_Figures}). 
Here, we should note that for the strong THz light ($\Omega_{\rm THz}\gg \omega_{\rm THz}$), the amplitude of each light source, $J_l(2\Omega_{{\rm THZ}}/\omega_{{\rm THz}})$, becomes an oscillating function of $l$. 
Therefore, the intensity of $n$-th order HHG becomes insensitive to $n$ after averaging out all the possible processes, which results in the plateau structure in HHG spectra. 
Furthermore, a semi-metallic state arises under the condition $2\Omega_{\rm THZ}\gtrsim E_{g}/\hbar$, because the direct resonant transition from the valence band to the conduction band (not taken into account in Fig.~\ref{fig:Schematic_Figures}) starts to be effective. 
The multiple-excitation picture implies that by tuning the NIR frequency, the HSG spectra can probe superpositions of various excitation paths.

In conclusion, we theoretically investigated the HSG in a semiconductor from the viewpoint of pump-probe spectroscopy with strong THz and weak NIR lights. 
We calculated the $\pm 2$nd HSG signal as a function of the NIR frequency and found that multiple resonant sub-peaks develop at $\omega_{\rm NIR} = E_{g}/{\hbar}+ k\omega_{\rm THz}$ ($k$ is an integer) as the THz light amplitude increases. 
The HSG signal shows non-monotonic behavior as a function of the amplitude of the THz light. 
Our analysis revealed that these features can be derived from the concept of a temporally changing band structure, and it is difficult to interpret them on the basis of the three-step model {or} the Bloch oscillation. 
The present analysis utilizing a unitary transformation indicates that the plateau structure in HHG spectra originates from a superposition of the various excitation paths constructed by multiple virtual light sources. 
This consideration may reveal the connection between HHG and other high-field phenomena, such as the above-threshold ionization, the dynamical Franz-Keldysh effect, and the coherent destruction of tunneling. 
The conclusions obtained in this paper would be useful even in regard to HHG in gaseous media, and they may be able to be used in experiments on both atomic and solid-state cases.

\section*{Acknowledgements}
The authors acknowledge to K. Tanaka for suggesting the topic treated in this paper.
The authors also acknowledge to T. Ikeda and A. Ishikawa for useful discussions.
This work was supported by JSPS KAKENHI Grant No. 19K14624.

\newpage

%%%%%%%%%% Merge with supplemental materials %%%%%%%%%%
\pagebreak
\widetext
\begin{center}
\textbf{\large Supplementary materials: Sub-band picture of high-harmonic generation in solids}
\end{center}
%%%%%%%%%% Merge with supplemental materials %%%%%%%%%%
%%%%%%%%%% Prefix a "S" to all equations, figures, tables and reset the counter %%%%%%%%%%
\setcounter{equation}{0}
\setcounter{figure}{0}
\setcounter{table}{0}
\setcounter{page}{1}
\makeatletter
\renewcommand{\theequation}{S\arabic{equation}}
\renewcommand{\thefigure}{S\arabic{figure}}
\renewcommand{\bibnumfmt}[1]{[S#1]}
\renewcommand{\citenumfont}[1]{S#1}
%%%%%%%%%% Prefix a "S" to all equations, figures, tables and reset the counter %%%%%%%%%%
\author{Tomohiro Tamaya${}^{*}$ and Takeo Kato${}$}
\affiliation{
${}$Institute for Solid State Physics, University of Tokyo, Kashiwa, 277-8581, Japan}

\date{\today}

\maketitle
\section{Derivation of the Hamiltonian}
Here, we will derive the Hamiltonian employed in our theory. 
Let us start from the microscopic Hamiltonian,
\begin{align}
H=\frac{1}{2m_0}\left(\bm{p}-\frac{e}{c} \sum_{\nu}\bm{A}_{\nu}(t)\right)^2
+\sum_{i}V(\bm{x}-\bm{R}_{i}),
\end{align}
where $m_{0}$ is the electron mass, $e${($<0$)} the electron charge, $\bm{p}$ the momentum of the bare electron, $c$ the velocity of light, $\bm{A}_{\nu}(t)$ ($\nu = {\rm NIR \ or \ THz}$) the vector potential of the incident electric fields, and $V(\bm{x}-\bm{R}_{i})$ the periodic core potential of atoms located at $\bm{R}_{i}$. 
Here, we will ignore the quasi-static energy $e^2 \bm{A}^{2}_{\nu}(t)^2/2m_{0}c^2$, which only shifts the total energy~\cite{Eberly1991S}. 
In this derivation, the incident light is assumed to have linearly polarized electric fields, described as
\begin{align}
\bm{A}_{\nu}(t)=\hat{\bm{x}}A_{\nu}(t)=\hat{\bm{x}}A_{\nu 0}\exp\left[-(t-t_{0})^2/T^2 \right] \cos \omega_\nu t,
\end{align}
where $A_{\nu 0}$ is the maximum amplitude of the $\nu$ incident electric field. 
For simplicity, we consider a basic two-dimensional covalent crystal with a simple lattice structure, which includes two atoms A and B in a unit cell keeping space-inversion symmetry. 
This assumption is equivalent to focusing {only} on the conduction and valence bands in a semiconductor. 
By employing the Coulomb gauge and supposing the tight-binding model {with} nearest-neighbor hopping of electrons, we can arrive at the following Hamiltonian in the second quantized form~\cite{Tamaya2016PRLS,Tamaya2016PRBRS,Tamaya2017ScienceS}
\begin{eqnarray}
H_{0}&=&\sum_{\bm{k}}[\gamma f\left(\bm{k}\right)a^{\dagger}_{\bm{k}}b_{\bm{k}} + \gamma f^{*}\left(\bm{k}\right)b^{\dagger}_{\bm{k}}a_{\bm{k}}], \label{Hamiltonian0S} \\
H_{I}&=&\hbar \sum_{\bm{k},\nu}[\Omega_{\nu}(\bm{k},t)a_{\bm{k}}^{\dagger}b_{\bm{k}} + \Omega^{*}_{\nu}(\bm{k},t)b_{\bm{k}}^{\dagger}a_{\bm{k}}].
\label{Hamiltonian1S}
\end{eqnarray}
Here, $\gamma$ is the transfer integral, $f(\bm{k})=\sum_{i}e^{i\bm{k}\cdot \bm{\delta}_{i}}=|f(\bm{k})|e^{i \theta_{f(\bm{k})}}$ {is the form factor}, $\bm{\delta}_{i}$ is the lattice vector, $a_{\bm{k}}$ ($b_{\bm{k}}$) is the annihilation operator of electrons with the wavenumber $\bm{k}$ on the sub lattice A (B), and $\Omega_{\nu}(\bm{k},t)$ is the Rabi frequency defined by
\begin{eqnarray}
\Omega_{\nu}(\bm{k},t)=\frac{e\hbar}{m_{0}c} \sum_{i}e^{i \bm{k}\cdot \bm{\delta}_{i}}\int d^2x \phi_{A}(\bm{x})\bm{A}_{\nu}(t)\cdot \bm{p} \phi_{B}(\bm{x}-\bm{\delta}_{i}),
\label{Rabi_frequencyS}
\end{eqnarray}
where $\phi_{A}(\bm{x})$ and $\phi_{B}(\bm{x})$ are the wave functions of electrons bound to atoms A and B. 
In the following formulation, we will ignore the $\bm{k}$ dependence of the Rabi frequency, which is {an approximation usually employed} in semiconductor physics~\cite{CardonaS}. 
The transformation into the band-structure picture can be performed by diagonalization of the single-particle part $H_{0}$ through the use of a unitary transformation defined as 
\begin{eqnarray}
a_{c\bm{k}}&=&\frac{1}{\sqrt{2}}[a_{\bm{k}}+e^{i\theta_{f(\bm{k})}}b_{\bm{k}}], \\
a_{v\bm{k}}&=&i\frac{1}{\sqrt{2}}[-a_{\bm{k}}+e^{i\theta_{f(\bm{k})}}b_{\bm{k}}].
\label{UnitaryS}
\end{eqnarray}
Substituting these expressions into Eqs.~(\ref{Hamiltonian0S}) and (\ref{Hamiltonian1S}), the Hamiltonian in the conduction-valence band is derived as $H=H_{0}+H_{I}$, where
\begin{eqnarray}
H_{0}&=&\sum_{\bm{k}}[E_{c}\left(\bm{k}\right)a^{\dagger}_{c\bm{k}}a_{c\bm{k}} + E_{v}\left(\bm{k}\right)a^{\dagger}_{v\bm{k}}a_{v\bm{k}}], \label{Hamiltonian200S}\\
H_{I}&=&\hbar \sum_{\bm{k},\nu}[\Omega_{\nu}(t)\cos \theta_{f(\bm{k})}(a_{c\bm{k}}^{\dagger}a_{c\bm{k}}-a_{v\bm{k}}^{\dagger}a_{v\bm{k}}) - \Omega_{\nu}(t)\sin \theta_{f(\bm{k})}(a_{v\bm{k}}^{\dagger}a_{c\bm{k}}+a_{c\bm{k}}^{\dagger}a_{v\bm{k}})].
\label{Hamiltonian210S}
\end{eqnarray}
Here, we suppose $\gamma |f(\bm{k})|= E_{c}(\bm{k})=-E_{v}(\bm{k})=E_{g}/2+\hbar^2 \bm{k}^2/2 m_{c}=E_{g}/2-\hbar^2 \bm{k}^2/2 m_{v} \equiv E(\bm k)${, where the isotropic band structures, i.e., $m_{c}=-m_{v}$, is assumed}. 
The first and second terms of Eq.~(\ref{Hamiltonian210S}) describe the intra and interband transitions of Bloch electrons, respectively. 
We also assume the relationship $\theta_{f(\bm{k})}=\theta_{\bm{k}}$, which can be justified in the honeycomb lattice structure near the energy gap (the $K$ points). 
Examples of this assumption can be found in discussions on graphene systems~\cite{MalicS,StrouckenS}. 
Although the general lattice structure will modify the $\theta_{\bm k}$-dependence in the matter-light interaction Hamiltonian, the overall features such as the plateau structure, semimetal characteristics, and two-color dynamics in high-harmonic generation in solids are expected to be unchanged.

\section{Derivation of the time evolution equations}
Here, we derive the time evolution equations of the wavefunction ${\psi}_{c\bm{k}}$ (${\psi}_{v\bm{k}}$) for conduction (valence) electrons. 
In the above Hamiltonian, the first term of Eq.~(\ref{Hamiltonian210S}) can be {absorbed} into Eq.~(\ref{Hamiltonian200S}). 
Thus, the Hamiltonian can be expressed in the form,
\begin{eqnarray}
H&=&\sum_{\bm{k}}[{\xi}_{c}(\bm{k}{,t})a^{\dagger}_{c\bm{k}}a_{c\bm{k}} + {\xi}_{v}(\bm{k}{,t})a^{\dagger}_{v\bm{k}}a_{v\bm{k}}]-\hbar\sum_{\bm{k},\nu}\Omega_{\nu}(t)\sin \theta_{\bm{k}}(a^{\dagger}_{v\bm{k}}a_{c\bm{k}}+a^{\dagger}_{c\bm{k}}a_{v\bm{k}})], 
\label{Hamiltonian3S}
\end{eqnarray}
where ${{\xi}_{c}(\bm{k},t)=-{\xi}_{v}(\bm{k},t)= E(\bm{k}) + \cos\theta_{\bm{k}} \sum_{\nu}\hbar \Omega_{\nu}(t)}$. 
Using this Hamiltonian, the time evolution equations for field operators $a_{c(v)\bm{k}}$ can be derived as
\begin{eqnarray}
i\hbar \dot{a}_{c\bm{k}}&=& [a_{c\bm{k}},H]={\xi}_{c}(\bm{k}{,t})a_{c\bm{k}}-\sum_{\nu}\hbar \Omega_{\nu}(t)\sin\theta_{\bm{k}}a_{v\bm{k}}, 
\label{Kinetic_Equations1S} \\
i\hbar \dot{a}_{v\bm{k}}&=& [a_{v\bm{k}},H]={\xi}_{{v}}(\bm{k}{,t})a_{v\bm{k}}-\sum_{\nu}\hbar \Omega_{\nu}(t)\sin\theta_{\bm{k}}a_{c\bm{k}}.
\label{Kinetic_Equations2S}
\end{eqnarray}
We assume that the system is initially in the ground state, i.e., in the state that all the covalent (conduction) states are occupied (unoccupied) by electrons. 
Accordingly, the time evolution is described as 
\begin{align}
\ket{\psi(t)} = \prod_{\bm k} \ket{\psi_{\bm k}(t)}
= \prod_{\bm k} \left[\psi_{c{\bm k}}(t) a^{\dagger}_{c{\bm k}}\ket{0} + \psi_{v{\bm k}}(t) a^{\dagger}_{v{\bm k}}\ket{0} \right],
\label{eq:wfS}
\end{align}
where $\ket{0}$ is the vacuum state. 
From the equations of motion, Eqs.~(\ref{Kinetic_Equations1S}) and (\ref{Kinetic_Equations2S}), we obtain the equations for $\psi_{c{\bm k}}(t)$ and $\psi_{v{\bm k}}(t)$:
\begin{eqnarray}
 i \hbar \dot{{\psi}}_{c\bm{k}}&=&{\xi}_{c}(\bm{k},t){\psi}_{c\bm{k}}- \sin\theta_{\bm{k}}\sum_{\nu}\hbar{\Omega_{\nu}(t)} {\psi}_{c\bm{k}}, \label{eq:kincS} \\
 i \hbar \dot{{\psi}}_{v\bm{k}}&=&{\xi}_{v}(\bm{k},t){\psi}_{v\bm{k}}-\sin\theta_{\bm{k}}\sum_{\nu}\hbar{\Omega_{\nu}(t)} {\psi}_{v\bm{k}}. \label{eq:kinvS}
\end{eqnarray}
{We note that these two equations for ${\psi}_{c\bm{k}}$ and ${\psi}_{v\bm{k}}$ have the same form as Eqs.~(\ref{Kinetic_Equations1S}) and (\ref{Kinetic_Equations2S}) for $a_{c\bm{k}}$ and $a_{v\bm{k}}$.}
The solutions of these equations provide the time evolution of ${\psi}_{c\bm{k}}$ and ${\psi}_{v\bm{k}}$. 
The temporal developments of the generated currents are derived from the definition, ${\cal J}{(t)}=-c\sum_{\nu}\langle {\partial H_{I}}/{\partial A_{\nu}(t)}\rangle \propto \sum_{\bm{k}}{\langle} \cos\theta_{\bm{k}} ({a}^{\dagger}_{c\bm{k}}{a}_{c\bm{k}}-{a}^{\dagger}_{v\bm{k}}{a}_{v\bm{k}}) -\sin\theta_{\bm{k}}({a}^{\dagger}_{v\bm{k}} {a}_{c\bm{k}}+{a}^{\dagger}_{c\bm{k}}{a}_{v\bm{k}}){\rangle}$. 
%By replacing $a_{c \bm k}$ and $a_{v \bm k}$ with ${\psi}_{c\bm{k}}$ and ${\psi}_{v\bm{k}}$, we obtain the generated current in the form, ${{\cal J}(t)}\propto \sum_{\bm{k}}[\cos\theta_{\bm{k}} (|{\psi}_{c\bm{k}}|^2-|{\psi}_{v\bm{k}}|^2) -\sin\theta_{\bm{k}}({\psi}^{*}_{v\bm{k}} {\psi}_{c\bm{k}}+{\psi}^{*}_{c\bm{k}}{\psi}_{v\bm{k}})]$.
For the wavefunction given in Eq.~(\ref{eq:wfS}), the average of the current is given by ${{\cal J}(t)} {=} \bra{\psi(t)} {\cal J} \ket{\psi(t)} \propto \sum_{\bm{k}}[\cos\theta_{\bm{k}} (|{\psi}_{c\bm{k}}|^2-|{\psi}_{v\bm{k}}|^2) -\sin\theta_{\bm{k}}({\psi}^{*}_{v\bm{k}} {\psi}_{c\bm{k}}+{\psi}^{*}_{c\bm{k}}{\psi}_{v\bm{k}})]$.

\section{Unitary Transformation of the Hamiltonian}

Here, we will perform a unitary transformation on the above Hamiltonian {to} obtain the rigid-band picture of the system. 
{For simplicity, we will consider only THz light by setting the Rabi frequencies as $\Omega_{\rm NIR}(t)=0$ and $\Omega_{\rm THz}(t)=\Omega \cos\omega t$. 
The Hamiltonian (\ref{Hamiltonian200S}) and (\ref{Hamiltonian210S}) is rewritten as $H=H_0+H_I$, where}
\begin{eqnarray}
H_{0}&=&\sum_{\bm{k}}E\left(\bm{k}\right)(a^{\dagger}_{c\bm{k}}a_{c\bm{k}} - a^{\dagger}_{v\bm{k}}a_{v\bm{k}}), \label{Hamiltonian20S}\\
H_{I}&=&\hbar {\Omega \cos\omega t} \sum_{\bm{k}} [\cos \theta_{\bm{k}}(a_{c\bm{k}}^{\dagger}a_{c\bm{k}}-a_{v\bm{k}}^{\dagger}a_{v\bm{k}}) - \sin \theta_{\bm{k}}(a_{v\bm{k}}^{\dagger}a_{c\bm{k}}+a_{c\bm{k}}^{\dagger}a_{v\bm{k}})].
\label{Hamiltonian21S}
\end{eqnarray}
By using the matrix form, the above Hamiltonian can be transformed into
\begin{eqnarray} 
H &=& \sum_{\bm k} 
( a^\dagger_{c{\bm k}} \ \ a^\dagger_{v{\bm k}} ) 
\left( \begin{array}{cc} 
E(\bm{k})+\hbar {\Omega}\cos \theta_{\bm{k}} \cos\omega t & -\hbar {\Omega}\sin \theta_{\bm{k}} \cos\omega t \\ -\hbar {\Omega}\sin \theta_{\bm{k}} \cos\omega t & -E(\bm{k})-\hbar {\Omega}\cos \theta_{\bm{k}} \cos\omega t
\end{array} \right)
\left( \begin{array}{c}
 a_{c{\bm k}} \\
 a_{v{\bm k}} \end{array} \right) \nonumber \\
 & \hspace{0mm}\equiv& \sum_{\bm k} 
( a^\dagger_{c{\bm k}} \ \ a^\dagger_{v{\bm k}} ) \tilde{H}_{\bm{k}}
\left( \begin{array}{c}
 a_{c{\bm k}} \\
 a_{v{\bm k}} \end{array} \right), \nonumber 
\end{eqnarray}
where $\tilde{H}_{\bm k}$ is a $2\times 2$ matrix whose components are real numbers.

For convenience, we will decompose the unitary transformation $U(\bm{k},t)$ defined in the main text into 
\begin{eqnarray}
U(\bm{k},t) &=& \left( \begin{array}{cc} 
e^{i\lambda}\sin(\frac{\theta_{\bm{k}}}{2}) & -e^{-i\lambda}\cos(\frac{\theta_{\bm{k}}}{2}) \\ e^{i\lambda}\cos(\frac{\theta_{\bm{k}}}{2}) & e^{-i\lambda}\sin(\frac{\theta_{\bm{k}}}{2})
\end{array} \right) \nonumber \\
&=& \left( \begin{array}{cc} 
\sin\frac{\theta_{\bm{k}}}{2} & \cos\frac{\theta_{\bm{k}}}{2} \\ \cos\frac{\theta_{\bm{k}}}{2} & -\sin\frac{\theta_{\bm{k}}}{2}
\end{array} \right) 
\left( \begin{array}{cc} 
e^{i\lambda{(t)}} & 0 \\ 0 & -e^{-i\lambda{(t)}}
\end{array} \right) \equiv U_{\theta_{\bm k}}U(t) \nonumber 
\label{Unitary2S}
\end{eqnarray}
Here, we define $\lambda(t)$ as a function of $t$ to be determined later. 
The unitary matrix $U_{\theta_{\bm{k}}}$ only depends on the wavenumber $\bm{k}$, while $U(t)$ depends on time and requires reconsideration of the unitary transformation based on the {Heisenberg} equation. 
Thus, we can easily perform a unitary transformation defined by the matrix $U_{\theta_{\bm{k}}}$ on the Hamiltonian and obtain, 
\begin{eqnarray}
H'&=&\sum_{\bm{k}}( s^\dagger_{c{\bm k}} \ \ s^\dagger_{v{\bm k}} ) U_{\theta_{\bm k}}^{-1} H_{\bm{k}} U_{\theta_{\bm k}} \left( \begin{array}{c}
 s_{c{\bm k}} \\
 s_{v{\bm k}} \end{array} \right) \nonumber \\
 &=&\sum_{\bm{k}}( s^\dagger_{c{\bm k}} \ \ s^\dagger_{v{\bm k}} ) \left( \begin{array}{cc} 
E(\bm{k})\cos \theta_{\bm{k}} + \hbar \Omega\cos \omega t & E(\bm{k})\sin \theta_{\bm{k}} \\ E(\bm{k})\sin \theta_{\bm{k}} & -E(\bm{k})\cos \theta_{\bm{k}} - \hbar \Omega\cos \omega t
\end{array} \right) \left( \begin{array}{c}
 s_{c{\bm k}} \\
 s_{v{\bm k}} \end{array} \right), \nonumber
\label{Hamiltonian_Unitary1S}
\end{eqnarray}
where the field operators $s_{c\bm{k}}$ and $s_{v\bm{k}}$ are the new basis for the Hamiltonian.

Next, let us focus on the time-dependent unitary transformation with $U(t)$, which corresponds to a gauge transformation modifying the time-dependent shift of the relative phase between the conduction and valence states. 
We start from the Heisenberg equation for the annihilation operators of the conduction and valence electrons, $s_{c \bm k}$ and $s_{v \bm k}$ (after a unitary transformation with $U_{\theta_{\bm k}}$):
\begin{eqnarray}
i \hbar \frac{d}{dt} \left( \begin{array}{c}
 s_{c{\bm k}} \\
 s_{v{\bm k}} \end{array} \right)&=&\left( \begin{array}{cc} 
E(\bm{k})\cos \theta_{\bm{k}} + \hbar \Omega\cos \omega t & E(\bm{k})\sin \theta_{\bm{k}} \\ E(\bm{k})\sin \theta_{\bm{k}} & -E(\bm{k})\cos \theta_{\bm{k}} - \hbar \Omega\cos \omega t
\end{array} \right) \left( \begin{array}{c}
 s_{c{\bm k}} \\
 s_{v{\bm k}} \end{array} \right) \nonumber \\
&\equiv& H'_{\bm k}\left( \begin{array}{c}
 s_{c{\bm k}} \\
 s_{v{\bm k}} \end{array} \right) \nonumber.
\label{Schrodinger00S}
\end{eqnarray}
By performing the unitary transformation $( s_{c{\bm k}}, s_{v{\bm k}} )^T=U(t) ( \alpha_{c{\bm k}}, \alpha_{v{\bm k}} )^T$, the above {Heiseberg} equation becomes
\begin{eqnarray}
& & i \hbar \frac{d}{dt}\left( \begin{array}{c}
 s_{c{\bm k}} \\
 s_{v{\bm k}} \end{array} \right)=i \hbar \dot{U}(t)\left( \begin{array}{c}
 \alpha_{c{\bm k}} \\
 \alpha_{v{\bm k}} \end{array} \right)+i \hbar U(t) \left( \begin{array}{c}
 \dot{\alpha}_{c{\bm k}} \\
 \dot{\alpha}_{v{\bm k}} \end{array} \right)=H_{\bm k}' U(t)\left( \begin{array}{c}
 \alpha_{c{\bm k}} \\
 \alpha_{v{\bm k}} \end{array} \right),
\label{Schrodinger2S} \nonumber \\
& & \Longleftrightarrow \quad 
i \hbar \frac{d}{dt} \left( \begin{array}{c}
 \alpha_{c{\bm k}} \\
 \alpha_{v{\bm k}} \end{array} \right)=(U^{-1}(t)H_{\bm k}' U(t)-i \hbar U^{-1}(t)\dot{U}(t))\left( \begin{array}{c}
 \alpha_{c{\bm k}} \\
 \alpha_{v{\bm k}} \end{array} \right) \equiv H_{\bm k}'' \left( \begin{array}{c}
 \alpha_{c{\bm k}} \\
 \alpha_{v{\bm k}} \end{array} \right).
\label{Schrodinger3S}
\end{eqnarray}
Therefore, the transformed Hamltonian, $H_{\bm k}''$, is given by
\begin{equation}
{H_{\bm k}'' = U^{-1}(t)H_{\bm k}' U(t)-i \hbar U^{-1}(t)\dot{U}(t).}
\end{equation}
The first and second terms in $H_{\bm k}''$ are calculated as
\begin{eqnarray}
U^{-1}(t)H_{\bm k}' U(t)=\left( \begin{array}{cc} 
E(\bm{k}) \cos \theta_{\bm{k}}+\Omega \cos \omega t & e^{-2 \lambda{(t)}} E(\bm{k})\sin \theta_{\bm{k}} \\ e^{2 \lambda{(t)}} E(\bm{k})\sin \theta_{\bm{k}} & -E(\bm{k}) \cos \theta_{\bm{k}} -\Omega \cos \omega t 
\end{array} \right)
\end{eqnarray}
and
\begin{eqnarray}
i \hbar U^{-1}(t)\dot{U}(t)= i \hbar \left( \begin{array}{cc} 
i\Omega \cos \omega t & 0 \\ 0 & -i\Omega \cos \omega t 
\end{array} \right).
\end{eqnarray}
Here, we {have defined} $\lambda (t)= -(\Omega/\omega)\sin \omega t$ {so that the time-dependent parts of the diagonal elements disappear in the transformed Hamiltonian}. 
Thus, the Hamiltonian of the system is described as
\begin{eqnarray}
H''_{\bm k}&=&\left( \begin{array}{cc} 
E(\bm{k}) \cos \theta_{\bm{k}} & e^{-2\lambda}E(\bm{k})\sin \theta_{\bm{k}} \\ e^{2\lambda}E(\bm{k})\sin \theta_{\bm{k}} & -E(\bm{k}) \cos \theta_{\bm{k}}
\end{array} \right) \nonumber \\
&=&\left( \begin{array}{cc} 
E(\bm{k}) \cos \theta_{\bm{k}} & E(\bm{k})\sin \theta_{\bm{k}}\sum_{n}J_{n}(\frac{2\Omega}{\omega})e^{i n \omega t} \\ E(\bm{k})\sin \theta_{\bm{k}}\sum_{n}J_{n}(\frac{2\Omega}{\omega})e^{-i n \omega t} & -E(\bm{k}) \cos \theta_{\bm{k}}
\end{array} \right), \nonumber
\end{eqnarray}
Here, we use the relationship $\exp[\pm i(2\Omega/\omega)\sin \omega t]=\sum_{n}J_{n}(2\Omega/\omega)e^{\pm i n \omega t}$. 
Thus, the Hamiltonian of the system can be expressed as
\begin{eqnarray}
H&=&\sum_{\bm{k}}( \alpha^\dagger_{c{\bm k}} \ \ \alpha^\dagger_{v{\bm k}} ) H''_{\bm{k}} \left( \begin{array}{c}
 \alpha_{c{\bm k}} \\
 \alpha_{v{\bm k}} \end{array} \right) \nonumber \\
&=&\sum_{\bm{k}}( \alpha^\dagger_{c{\bm k}} \ \ \alpha^\dagger_{v{\bm k}} ) \left( \begin{array}{cc} 
E(\bm{k}) \cos \theta_{\bm{k}} & E(\bm{k})\sin \theta_{\bm{k}}\sum_{n}J_{n}(\frac{2\Omega}{\omega})e^{i n \omega t} \\ E(\bm{k})\sin \theta_{\bm{k}}\sum_{n}J_{n}(\frac{2\Omega}{\omega})e^{-i n \omega t} & -E(\bm{k}) \cos \theta_{\bm{k}}
\end{array} \right) \left( \begin{array}{c}
 \alpha_{c{\bm k}} \\
 \alpha_{v{\bm k}} \end{array} \right). \nonumber
 \label{eq:Hamiltonian2S}
\end{eqnarray}
By performing the same unitary transformation on ${\cal J}(t)$, we obtain a new expression of the generated current in the rigid-band picture. 
By using the first unitary transformation $( a_{c{\bm k}}, a_{c{\bm k}})^T = U_{\theta_{\bm k}} ( s_{c{\bm k}}, s_{c{\bm k}})^T$, the current is transformed from, ${\cal J}{(t)} \propto \sum_{\bm{k}}{\langle}\cos\theta_{\bm{k}} ({a}^{\dagger}_{c\bm{k}}{a}_{c\bm{k}}-{a}^{\dagger}_{v\bm{k}}{a}_{v\bm{k}}) -\sin\theta_{\bm{k}}({a}^{\dagger}_{v\bm{k}} {a}_{c\bm{k}}+{a}^{\dagger}_{c\bm{k}}{a}_{v\bm{k}}){\rangle}$, to, ${{\cal J}{(t)}}\propto \sum_{\bm{k}}{\langle}{s}^{\dagger}_{c\bm{k}}{s}_{c\bm{k}}-{s}^{\dagger}_{v\bm{k}}{s}_{v\bm{k}}{\rangle}$. 
%Subsequently, by using the time-dependent unitary transformation $( s_{c{\bm k}}, s_{v{\bm k}} )^T=U(t) ( \alpha_{c{\bm k}}, \alpha_{v{\bm k}} )^T$, we obtain ${\cal J}\del{(t)}$. 
Subsequently, by using the time-dependent unitary transformation $( s_{c{\bm k}}, s_{v{\bm k}} )^T=U(t) ( \alpha_{c{\bm k}}, \alpha_{v{\bm k}} )^T$, we obtain ${\cal J}{(t)} \propto \sum_{\bm{k}}{\langle}\alpha^{\dagger}_{c\bm{k}} \alpha_{c\bm{k}}-\alpha^{\dagger}_{v\bm{k}} \alpha_{v\bm{k}}{\rangle}$.
By expressing the wavefunction by the last basis as $\ket{\psi(t)} = \prod_{\bm k}[\phi_{c{\bm k}}(t) \alpha_{c{\bm k}}^\dagger \ket{0} +\phi_{v{\bm k}}(t)\alpha_{v{\bm k}}^\dagger \ket{0}]$, we obtain ${\cal J}{(t)} \propto \sum_{\bm{k}}[|\phi_{c\bm{k}}|^2-|\phi_{v\bm{k}}|^2]$ {in the Schr\"odinger picture}.
%By replacing $s_{c \bm{k}}$ and $s_{v \bm{k}}$ with $\phi_{c \bm{k}}$ and $\phi_{v \bm{k}}$, we {finally} arrive at the expression of the generated current as ${\cal J}(t)=\sum_{\bm{k}}[|\phi_{c\bm{k}}|^2-|\phi_{v\bm{k}}|^2]$. 
%The time evolution of $\phi_{c \bm{k}}$ and $\phi_{v \bm{k}}$ can be calculated from kinetic equations derived from the above Hamiltonian.
We note that the time evolution of $\phi_{c \bm{k}}$ and $\phi_{v \bm{k}}$ has the same form as the Heisenberg equation for $\alpha_{c{\bm k}}$ and $\alpha_{v{\bm k}}$ derived from the transformed Hamiltonian.


\begin{thebibliography}{10}
\bibitem{footnote}
We also expect the $n$th-order HHG to show the nonmonotonic behavior as a function of the field intensity, because the weight of the level plays important role in constructing plateau structures.
\end{thebibliography}
\end{document}